# Improved IKE Protocol Design Based On PKI/ECC

Pak Song-Ho, Pak Myong-Suk, Jang Chung-Hyok

**Kim Il Sung** University,

Pyongyang, DPR of Korea

Abstract  This Paper proposes an ECDH key exchange method and an ECsig Digital Signature Authentication method based on $E(F_{2^m})$ group with Koblits curve, man-in-the-middle attack prevention method for SA payload and initiator identification payload to design high intensity IKE that can be implemented in portable devices.

Keywords  IPSec, ECC, IKE, SA, Koblitz

## 1. Introduction

IPSec framework is a set of open standards developed by the IETF, which uses IKE(Internet Key Exchange)protocol to create and exchange key between IPSec peers and authenticate remote peer.

IKE is a two-phase, multimode protocol that offers three authentication methods such as preshared key, public key signature and public key encryption. [1]

IKE 1st phase Security Association has various problems at raising security performance.

IPSec uses DH key exchange and RSA signature during IKE 1st phase, therefore it has slow modular exponentiation calculation and attack algorithms that attacker can find secure key.

To design high intensity IKE, Previous works[1] already defined $E(F_{2^m})$ group for IKE, but the elliptic curves over $E(F_{2^{155}})$ group and $E(F_{2^{185}})$ group have non prime extension degree( $m$ ), therefore security performance is not guarantee.

And previous works[2,3] defined elliptic curve group $E(F_p)$, but the operation over prime field $F_p$ is less effective than the operation over binary field $F_{2^m}$ in implementing hardware, so it can't be used to implement portable device.

Next, all of the man-in-the middle attack prevention methods excepting initiator identification protection method have been proposed in the case of using preshared key authentication mode, because two IPSec peers have already shared key previously.

But because IKE 1st phase authenticated with signature has not the preshared key between IPSec peers, it can not use above methods.

Therefore, in this case, we must protect initiator identification from attacker.

Paper proposes an ECDH key exchange method and an ECsig Digital Signature Authentication method based on $E(F_{2^m})$ group with Koblits curve, man-in-the-middle attack prevention method for SA payload and initiator identification payload .

## 2. IKE Design method with ECDH key exchange and ECSig signature

1) Selection of $E(F_{2^m})$ group with Koblitz curve and the DH group identifier Definition in IKE 1$^{st}$ phase security association.

Elliptic curve group $E(F_p)$ defined over prime field $F_p$ and elliptic curve group $E(F_{2^m})$ defined over binary field $F_{2^m}$ are usually used in elliptic curve cryptography,

Because operation over $F_{2^m}$ is more effective than operation over $F_p$ in implementing hardware, EC group over $F_{2^m}$ is widely used in implementing commercially.

Therefore, in this paper we select five Koblitz curves defined over $F_{2^m}$ among elliptic curves recommended by NIST and implements these curves in IPSec.

Koblitz curves used in this paper are as follows.

**K-163**: $a = 1, b = 1, h = 2, f(x) = x^{163} + x^7 + x^6 + x^3 + 1$

$Np$ = 0x 00000004 00000000 00000000 00020108 A2E0CC0D 99F8A5EF

**K-233**: $a = 0, b = 1, h = 4, f(x) = x^{233} + x^{74} + 1$

$Np$ = 0x 00000080 00000000 00000000 00000000 00069D5B B915BCD4 6EFB1AD5 F173ABDF

**K-283**: $a = 0, b = 1, h = 4, f(x) = x^{283} + x^{12} + x^7 + x^5 + 1$

$Np$ = 0x 01FFFFFF FFFFFFFF FFFFFFFF FFFFFFFF FFFFE9AE 2ED07577 265DFF7F 265DFF7F 94451E06 1E163C61

**K-409**: $a = 0, b = 1, h = 4, f(x) = x^{409} + x^{87} + 1$

$Np$ = 0x 007FFFFF FFFFFFFF FFFFFFFF FFFFFFFF FFFFFFFF FFFFFFFF FFFFFE5F 83B2D4EA 20400EC4 557D5ED3 E3E7CA5B 4B5C83B8 E01E5FCF

**K-571**: $a = 0, b = 1, h = 4, f(x) = x^{571} + x^{10} + x^5 + x^2 + 1$

$Np$ = 0x 02000000 00000000 00000000 00000000 00000000 00000000 00000000 00000000 00000000 131850E1 F19A63E4 B391A8DB 917F4138 B630D84B E5D63938 1E91DEB4 5CFE778F 637C1001

Where, $a$ and $b$ are coefficients, $f(x)$ is reduction, $N_P$ is group order(prime) of basic point($P$), $h$ is co-factor and $N_P \times h$ is the number of points over Koblitz curve.

To indicate groups for ECDH key exchange proposed in this paper, optional DH group identifiers except that must or should be used in default IKE are used.

$E(F_{2^{163}})$ group : DH group identifier 5

$E(F_{2^{233}})$ group : DH group identifier 15

$E(F_{2^{283}})$ group : DH group identifier 16

$E(F_{2^{409}})$ group : DH group identifier 17

$E(F_{2^{571}})$ group : DH group identifier 18

Origenal DH group identifiers 5 indicates 1536bit modp group and DH group identifier 15 ~ 18 indicate 3072bit modp group, 4096bit modp group, 6144bit modp group, 8192bit modp group respectively.

2) Main mode of IKE 1st phase security association with ECDH key exchange and ECSig signature

Figure 1 shows main mode of IKE 1st phase security association with ECDH key exchange and ECSig signature.

```
    initiator (i)              responder(r)
    -----------                -----------
①  HDRi, SAi          →
②                     ←   HDRr, SAr
③  HDRi, KEi, Ni      →
④                     ←   HDRr, KEr, Nr
⑤  HDRi*, IDii, [ CERT, ] SIG_I  →
⑥                     ←   HDRr*, IDir, [ CERT, ]  SIG_R
```

Figure 1. main mode of IKE 1st phase with ECDH key exchange and ECSig signature.

Here, SAi and SAr indicate SA payload of initiator and responder, KE is key exchange payload which contains the ECDH public key (for example, in the case of initiator, **KEi**= Ki×P), N is nonce payload, ID is identification payload (for example, in the case of initiator, IDii), CERT is the ECC certificate payload of initiator or responder, SIG_I and SIG_R are the ECSig signature payloads of initiator and responder.

－computing keying material

When exchanging ECDH public key, the following keying material are used and the value SKEYID is computed separately for each authentication method.

For signatures: SKEYID = prf(Ni_b | Nr_b, Ki×Kr×P)

Here, Ki and Kr are secure keys of initiator and responder, P is basic point of elliptic curve group $E(F_{2^m})$ and the meaning of rest payloads is the same as default.

For public key encryption: SKEYID = prf(hash(Ni_b | Nr_b), CKY-I | CKY-R)

For pre-shared keys: SKEYID = prf(pre-shared-key, Ni_b |Nr_b)

The result of either Main Mode or Aggressive Mode is three groups of authenticated keying material and these materials are as follows:

SKEYID_d = prf(SKEYID, Ki×Kr×P | CKY-I | CKY-R | 0)

SKEYID_a = prf(SKEYID, SKEYID_d | Ki×Kr×P | CKY-I | CKY-R | 1)

SKEYID_e = prf(SKEYID, SKEYID_a | Ki×Kr×P | CKY-I | CKY-R | 2)

The values of 0, 1, and 2 above are represented by a single octet. The key used for encryption is derived from SKEYID_e.

To authenticate exchange, the initiator of the protocol generates HASH_I and the responder generates HASH_R.

Where:

HASH_I = prf(SKEYID, Ki×P | Kr×P | CKY-I | CKY-R | SAi_b | IDii_b )

HASH_R = prf(SKEYID, Kr×P | Ki×P | CKY-R | CKY-I | SAi_b | IDir_b

For authentication with digital signatures, HASH_I and HASH_R are signed and verified; for authentication with either public key encryption or pre-shared keys, HASH_I and HASH_R directly authenticate the exchange.

prf(key, msg) is the keyed pseudo-random function and this function is used both for key derivations and for authentication.

－ using $E(F_{2^m})$ group with Koblitz curve in IKE $1^{st}$ phase security association

Before starting IKE $1^{st}$ phase security association, each initiator and responder has own ECC certificate from certificate authority (CA). At that time, two IPsec peers know DH group identifier number to use from public key length of ECC public key certificate issued by CA(two peers must have peer's ECC public key certificate)

## 3. Man-in-the-middle attack prevention method based on PKI/ECC in IKE $1^{st}$ phase security association

When using main mode with ECDH key exchange and ECsig signature, IKE $1^{st}$ phase security association with the confidentiality of SA payload and initiator identification protection is as follows.

    initiator                                responder

    ①  HDRi, {SAi}pr   →

② ← HDRr, {SAr}pi  
③ HDRi, KEi, Ni →  
④ ← HDRr, KEr,Nr,[IDir,[CERT,],SIG_R]$_{ECDH}$  
⑤ HDRi*, IDii,[CERT,]SIG_I →

Figure 2. Man-in-the-middle attack prevention method for main mode

Where, pr and pi indicate ECC public key of responder and initiator.

{ }x indicates that encrypts messages within the { } form with public key(x) using ECC.

[ ]$_{ECDH}$ shows that encrypts messages within the [ ] form with ECDH shared key value using secret key encryption algorithm.

When creating SIG_I and SIG_R, improved IKE uses the following formula.

HASH_I=prf(SKEYID, Ki×P|Kr×P|CKY-I|CKY-R|SAi_b|SAr_b | IDii_b )

HASH_R=prf(SKEYID, Kr×P|Ki×P|CKY-R|CKY-I|SAr_b|SAi_b|IDir_b

Above HASH formula shows integrity for SAi, SAr, IDii and IDir payload.

## 4. Character Evaluation and Conclusion

The table 1 and table 2 show comparison between previous IKE protocol and proposed method in this paper.

Table 1. Comparison of security performance, computing number, message exchange number, applying number

| Protocol name | Key exchange mode and computing number | Signature method and computing number | HASH use and computing number | message exchange number | Public key Cryptography applying number |
|---|---|---|---|---|---|
| IKE (IKE 1st phase main mode) | DH/ECDH(4) | RSASig/ ECSig (2) | default(2) | 6 | × |
| Improved IKE (IKE 1st phase main mode) | DH/ECDH(4) | RSASig/ ECSig (2) | improved(2) | 5 | ECC(2) |

Table 2. Comparison of security performance and applying number

| Protocol name | key use in secret key algorithm and appling number | identification protection | DoS attack prevention | SA payload protection | EC group used | Application device |
|---|---|---|---|---|---|---|
| IKE (IKE 1st phase main mode) | Same key use in 5,6 step(2) | ○ | × | × | $E(F_p)$ | desktop |
| Improved IKE (IKE 1st phase main mode) | different key use in 3, 4 step(2) | ○ | ○ | ○ | $E(F_p)$, $E(F_{2^m})$ | desktop, portable |

Where , × indicates no support , ○ indicates support.

As showed at this table, we know that improved IKE protocol has higher security performance and is suitable to implement in portable device than present IKE protocol.